\documentclass[aps,pra,twocolumn,groupedaddress,showpacs,amsmath,amssymb,amsfonts]{revtex4}

\usepackage{graphicx}

\usepackage{color}

\begin{document}

\title{Superfluid-Insulator Transitions in
Attractive Bose-Hubbard Model with Three-Body Constraint}

\author{Yu-Wen Lee}
\email{ywlee@thu.edu.tw} %
\affiliation{Department of Physics, Tunghai University, Taichung
40704, Taiwan}

\author{Min-Fong Yang}
\affiliation{Department of Physics, Tunghai University, Taichung
40704, Taiwan}

\date{\today}

\begin{abstract}
By means of the method of the effective potential, the phase
transitions from the Mott insulating state to either the atomic or
the dimer superfluid state in the three-body constrained
attractive Bose lattice gas are analyzed. Due to the appearance of
the Feshbach resonance coupling between the two kinds of order
parameters in the derived effective potential function, it is
found that the continuous Mott insulator-to-superfluid transitions
can be preempted by first-order ones. Since the employed approach
can provide accurate predictions of phase boundaries in the strong
coupling limit, where the dimer superfluid phase can emerge, our
work hence sheds light on the search of this novel phase in real
ultracold Bose gases in optical lattices.
\end{abstract}

\pacs{%
67.85.Hj,         
11.15.Me,         
64.70.Tg}         

\maketitle


The impressive developments on the manipulation of ultracold gases
in optical lattices provides one of the best environments in the
search for exotic quantum phases~\cite{BEC_reviews}. Because
unprecedented control over microscopic model parameters can be
achieved, it is possible to explore parameter regimes which are
not available in other analogous condensed matter systems. The
remarkable experimental demonstration of the superfluid to Mott
insulator transition in ultracold lattice bosons~\cite{BEC_exps}
has paved the way for investigating other strongly correlated
phases in similar setups. For instance, the search for novel and
unconventional quantum phases in the mixtures of ultracold atoms
obeying either the same or different statistics has attracted
considerable attention~\cite{BEC_reviews}.

It was recently suggested that intriguing quantum critical
behaviors can occur in attractive bosonic lattice gases with
three-body on-site constraint~\cite{Diehl09-1,Diehl09-2}. The
system is described by the Bose-Hubbard model with a three-body
constraint $a_i^{\dag\,3}\equiv 0$,
\begin{equation}
H = - t \sum_{\langle i,j\rangle} a_{i}^{\dagger }a_{j} %
    + \frac{U}{2} \sum_{i} n_{i}(n_{i} -1) - \mu \sum_{i} n_{i} \; ,
\label{eq:H}
\end{equation}
Here, $a_i (a_i^\dag)$ is the bosonic annihilation (creation)
operator at site $i$, $t$ is the hopping matrix element, $U<0$ the
on-site two-body attraction, and $\mu$ the chemical potential. The
convention $\langle i,j\rangle$ signifies a sum over
nearest-neighbor sites $i$ and $j$. The on-site constraint can
arise naturally due to large three-body loss
processes~\cite{Daley09,Roncaglia10}, and it stabilizes the
attractive bosonic system against collapse. Therefore, besides the
conventional atomic superfluid state (ASF) with non-vanishing
order parameters $\left\langle a \right\rangle \neq 0$ and
$\left\langle a^{2} \right\rangle \neq0 $ appearing in the
weakly-interacting limit, a dimer superfluid phase (DSF) with
vanishing atomic order parameter ($\left\langle a \right\rangle =
0$) but nonzero pairing correlation ($\left\langle a^{2}
\right\rangle \neq 0$) can be realized for sufficiently strong
attraction~\cite{Daley09}. It was shown in
Refs.~\cite{Diehl09-1,Diehl09-2} that this model provides a simple
realization of the physics of Ising quantum transition together
with the Coleman-Weinberg mechanism~\cite{Coleman-Weinberg}
without resorting to the Feshbach-resonant
mechanism~\cite{Radzihovsky04,Romans04,1QPT}. While the nature
around the ASF-DSF transition has been discussed, the detailed
physics of the Mott insulator (MI) to superfluid (either ASF or
DSF) transitions is not addressed in
Refs.~\cite{Diehl09-1,Diehl09-2}.

In the present work, we focus our attention on the MI-ASF and the
MI-DSF transitions in this three-body constrained attractive Bose
lattice gas.
Since there is no hopping term for dimers in the model of
Eq.~\eqref{eq:H}, usual strong-coupling theory based on simple
mean-field decoupling is not appropriate for our purpose, because
it fails to describe the DSF phase.
Instead, the method of the effective potential developed in
Refs.~\cite{dosSantos09,Bradlyn09} is employed, which has been
applied successfully to the usual Bose-Hubbard model with
\emph{repulsive} interaction. Accurate analytical results for the
phase boundaries of Mott insulator-to-superfluid transitions have
been obtained when the contributions from higher orders in the
hopping term are included
systematically~\cite{Teichmann09,Eckardt09}.
Here, we generalize their approach to the case with \emph{two
order parameters}. For the attractive bosonic lattice gases under
consideration, we show that the effective potential function has
the same form as the mean-field Ginzburg-Landau theory of resonant
Bose gases~\cite{Radzihovsky04,Romans04,1QPT}. Following similar
discussions in Ref.~\cite{1QPT}, it is found that the MI-DSF
transitions are always second-order, while the MI-ASF transitions
can be either second-order or first-order. The tricritical points
on the MI-ASF transition lines and the critical end points of the
MI-DSF phase boundaries are determined within the present
approach. From our results, it is shown that the DSF phase exists
only in a narrow region of chemical potential $\mu/|U|$ for small
hopping parameters $t/|U|$. Since the approach used in the present
work can be considered as a kind of strong-coupling expansion, it
is expected to provide accurate phase boundaries of the MI phases
for the transitions occurring at small values of $t/|U|$. Our
results hence provides a useful guide to the experimental search
of the DSF phase and the associated quantum phase transitions in
ultracold Bose gases in optical lattices.


To derive the effective potential with two order parameters for
the model in Eq.~\eqref{eq:H}, we begin by adding two symmetry
breaking source terms to the Hamiltonian, which are spatially and
temporally global:
$\tilde{H} = H_0 + V$ with $H_0$ being the single-site
zero-hopping contribution in Eq.~\eqref{eq:H} and
$V = - t \sum_{\langle i,j\rangle} a_{i}^{\dagger }a_{j} %
      + \sum_i (\chi^* a_i + \eta^* a_i^2 + {\rm H.c.})$.
It is easy to see that the exact ground state for the on-site part
$H_0$ is the $n=0$ MI state (i.e., the empty state) for the
chemical potential $\mu < -|U|/2$, while it becomes the $n=2$ MI
state (i.e., the completely filled state) for $\mu > -|U|/2$.
By treating $V$ as a perturbation and following the specific
adaption of high-order many-body perturbation theory proposed in
Refs.~\cite{dosSantos09,Bradlyn09},
the free energy (or the ground state energy at the $T=0$ limit
under consideration) in the MI phases for the modified Hamiltonian
$\tilde{H}$ 
can be calculated as a double power series in both the hopping
parameter $t$ and the source fields $\chi$, $\chi^*$ and $\eta$,
$\eta^*$. Up to the forth power in source fields, the general
expression of free energy per site $f$ takes the following form
\begin{align}
f \simeq & f_0 + r_1 |\chi|^2 + \frac{u_1}{2} |\chi|^4
  + r_2 |\eta|^2 + \frac{u_2}{2} |\eta|^4 \nonumber\\
& - \frac{\lambda}{2} (\eta^* \chi^2 + {\rm c.c.}) %
  + u_{12}|\chi|^2 |\eta|^2 \; , \label{eq:f}
\end{align}
where $f_0$ corresponds to the ground state energy density in the
absence of the source fields. We point out that our expression of
$f$ does respect the symmetry for the U(1) phase transformation
$\chi\rightarrow\chi\;e^{i\theta}$ and
$\eta\rightarrow\eta\;e^{2i\theta}$, which can be understood from
the form of the modified Hamiltonian $\tilde{H}$.

The coefficients in the expansion of free energy density $f$ in
Eq.~\eqref{eq:f} can be determined perturbatively in hopping
parameter $t$. In the following, energy unit is set to be $|U|$
and we define the dimensionless parameters $\bar{t} = t/|U|$ and
$\bar{\mu} = \mu/|U|$ for convenience. Up to leading order in
$\bar{t}$, we find that, for the $n=0$ MI state with
$\bar{\mu}<-1/2$,
\begin{align}
r_1& = \frac{1}{\bar{\mu}} %
       \left( 1 - \frac{z\bar{t}}{\bar{\mu}} %
       + \frac{z^2\bar{t}^2}{\bar{\mu}^2} \right) \; , \nonumber\\
u_1& = -\frac{2}{\bar{\mu}^3(2\bar{\mu}+1)} %
       \left( 1 - \frac{4z\bar{t}}{\bar{\mu}} \right) \; , \nonumber\\
r_2& = \frac{2}{2\bar{\mu}+1} %
       \left[ 1 + \frac{2z\bar{t}^2}{\bar{\mu}(2\bar{\mu}+1)} \right] \; , \label{eq:coef-0MI}\\
u_2& = -\frac{8}{(2\bar{\mu}+1)^3} \left[ 1 + O(\bar{t}^2) \right] \; , \nonumber\\
\lambda& = -\frac{4}{\bar{\mu}(2\bar{\mu}+1)} %
       \left( 1 + \frac{2z\bar{t}}{\bar{\mu}} \right) \; , \nonumber\\
u_{12}& = -\frac{2(3\bar{\mu}+1)}{\bar{\mu}^2(2\bar{\mu}+1)^2} %
       \left[ 1 - \frac{44\bar{\mu}^2+6\bar{\mu}-2}{\bar{\mu}(3\bar{\mu}+1)^2} z\bar{t} \right] \; ;\nonumber
\end{align}
while, for the $n=2$ MI state with $\bar{\mu}>-1/2$,
\begin{align}
r_1& = -\frac{2}{\bar{\mu}+1} %
       \left[ 1 + \frac{2z\bar{t}}{\bar{\mu}+1} %
       + \frac{4z^2\bar{t}^2}{(\bar{\mu}+1)^2} \right] \; , \nonumber\\
u_1& = \frac{4(3\bar{\mu}+1)}{(\bar{\mu}+1)^3(2\bar{\mu}+1)} %
       \left( 1 + \frac{8z\bar{t}}{3\bar{\mu}+1} \right) \; , \nonumber\\
r_2& = -\frac{2}{2\bar{\mu}+1} %
       \left[ 1 + \frac{2z\bar{t}^2}{(\bar{\mu}+1)(2\bar{\mu}+1)} \right] \; , \label{eq:coef-2MI}\\
u_2& = \frac{8}{(2\bar{\mu}+1)^3} \left[ 1 + O(\bar{t}^2) \right] \; , \nonumber\\
\lambda& = -\frac{4}{(\bar{\mu}+1)(2\bar{\mu}+1)} %
       \left( 1 - \frac{6z\bar{t}}{\bar{\mu}+1} \right) \; , \nonumber\\
u_{12}& = \frac{4(3\bar{\mu}+2)}{(\bar{\mu}+1)^2(2\bar{\mu}+1)^2} %
       \left[ 1 - \frac{72\bar{\mu}^2+96\bar{\mu}+32}{(\bar{\mu}+1)(3\bar{\mu}+2)^2} z\bar{t} \right] \; . \nonumber %
\end{align}
Here $z$ is the coordination number of the underlying lattices.
Form the free energy density $f$, the order parameters of the
atomic condensate $\phi_{\rm a} \equiv \langle a \rangle$ and the
molecular (or dimer) condensate $\phi_{\rm m} \equiv \langle a^2
\rangle$ can be obtained by the first derivative of $f$ with
respect to corresponding external sources. That is, $\phi_{\rm a}
= \partial f / \partial \chi^*$ and $\phi_{\rm m} = \partial f /
\partial \eta^*$, respectively. We note that, due to the mixing
terms in $f$ with coefficients $\lambda$ and $u_{12}$, there
exists a nontrivial relation between the two order parameters
$\phi_{\rm a}$ and $\phi_{\rm m}$.

The effective potential in terms of these order parameters is then
derived by performing the Legendre transformation on the free
energy density $f$,
$\Gamma(\phi_{\rm a}, \phi_{\rm a}^*, \phi_{\rm m}, \phi_{\rm m}^*)  %
= f - \chi^* \phi_{\rm a} - \chi \phi_{\rm a}^*  %
  - \eta^* \phi_{\rm m} - \eta \phi_{\rm m}^*$,
which can be used to determine the phase boundaries of the Mott
insulator-to-superfluid transition and their nature as shown
below. From Eq.~\eqref{eq:f}, the Ginzburg-Landau expansion of the
effective potential as a power series of the order parameter
variables can be obtained:
\begin{align}
& \Gamma(\phi_{\rm a}, \phi_{\rm a}^*, \phi_{\rm m}, \phi_{\rm m}^*) \nonumber\\
& \simeq  f_0 + m_1 |\phi_{\rm a}|^2 + \frac{g_1}{2} |\phi_{\rm a}|^4 %
   + m_2 |\phi_{\rm m}|^2 + \frac{g_2}{2} |\phi_{\rm m}|^4 \nonumber\\
&\quad  - \frac{\alpha}{2} (\phi_{\rm m}^* \phi_{\rm a}^2 + {\rm c.c.}) %
   + g_{12} |\phi_{\rm a}|^2 |\phi_{\rm m}|^2 \; , \label{eq:Gamma}
\end{align}
with the coefficients given by
$m_1 = -1/r_1$, $g_1 = u_1/r_1^4 + 3\lambda^2/2r_1^4 r_2$, $m_2 =
-1/r_2$, $g_2 = u_2/r_2^4$, $\alpha = \lambda/r_1^2 r_2$, and
$g_{12} = 3\lambda^2/r_1^3 r_2^2 - 3u_{12}/r_1^2 r_2^2$.
An interesting feature of the two-component Landau theory in
Eq.~\eqref{eq:Gamma} is the appearance of the Feshbach resonance
coupling with coefficient $\alpha$. The existence of such a term
has important consequences on the Ising quantum phase transition
in the boson Feshbach resonance
problems~\cite{Radzihovsky04,Romans04}. Starting from a specific
mean-field ansatz for the DSF state, the analogy between the
present problem and the usual boson Feshbach resonance model was
also noticed and explored in some detail in Ref.~\cite{Diehl09-2}.
Here we approach this issue from the MI states, and focus our
attention on its implication on the MI-ASF and the MI-DSF
transitions.

\begin{figure}[tb]
\includegraphics[width=0.9\columnwidth,height=0.7\columnwidth]{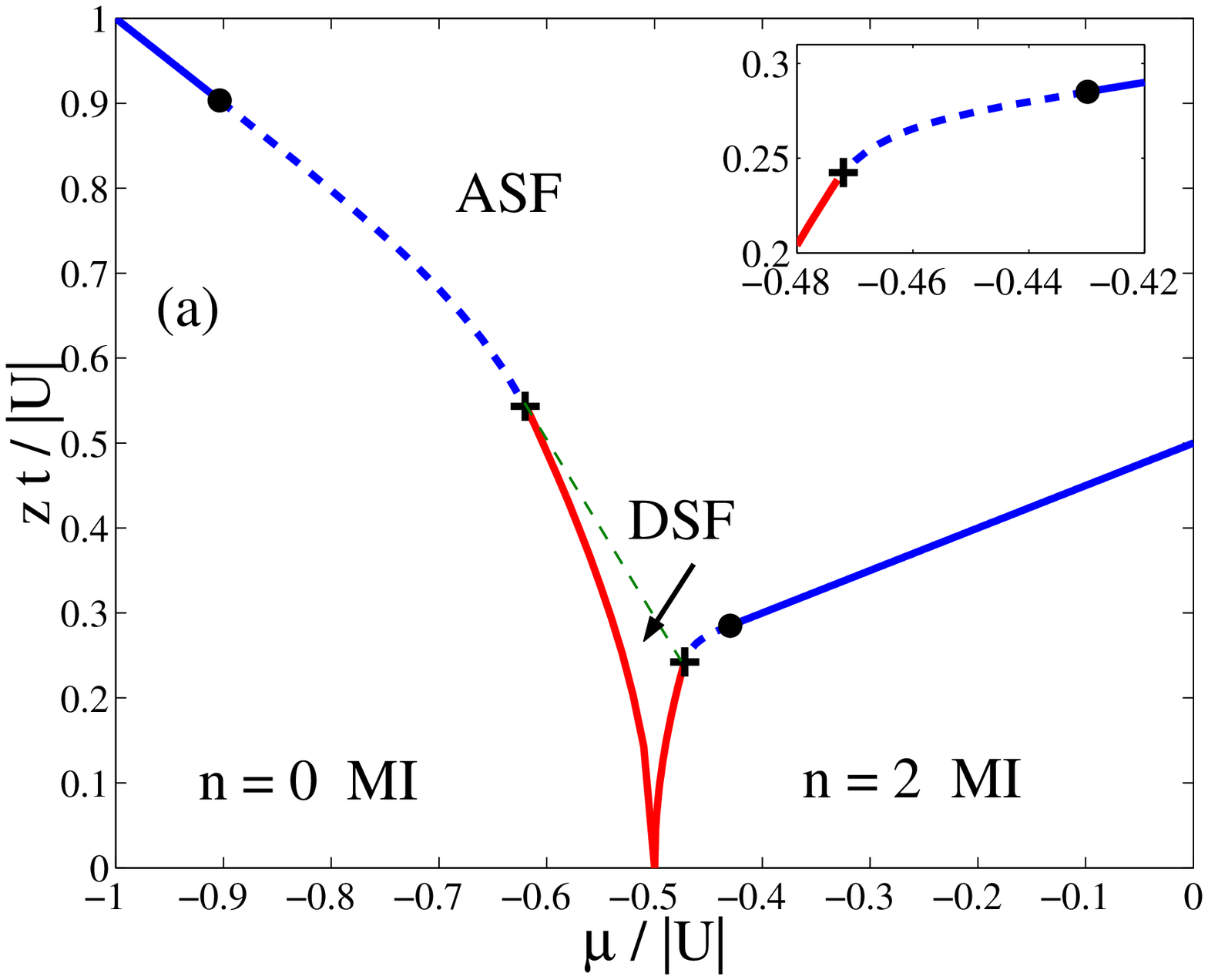}
\includegraphics[width=0.9\columnwidth,height=0.7\columnwidth]{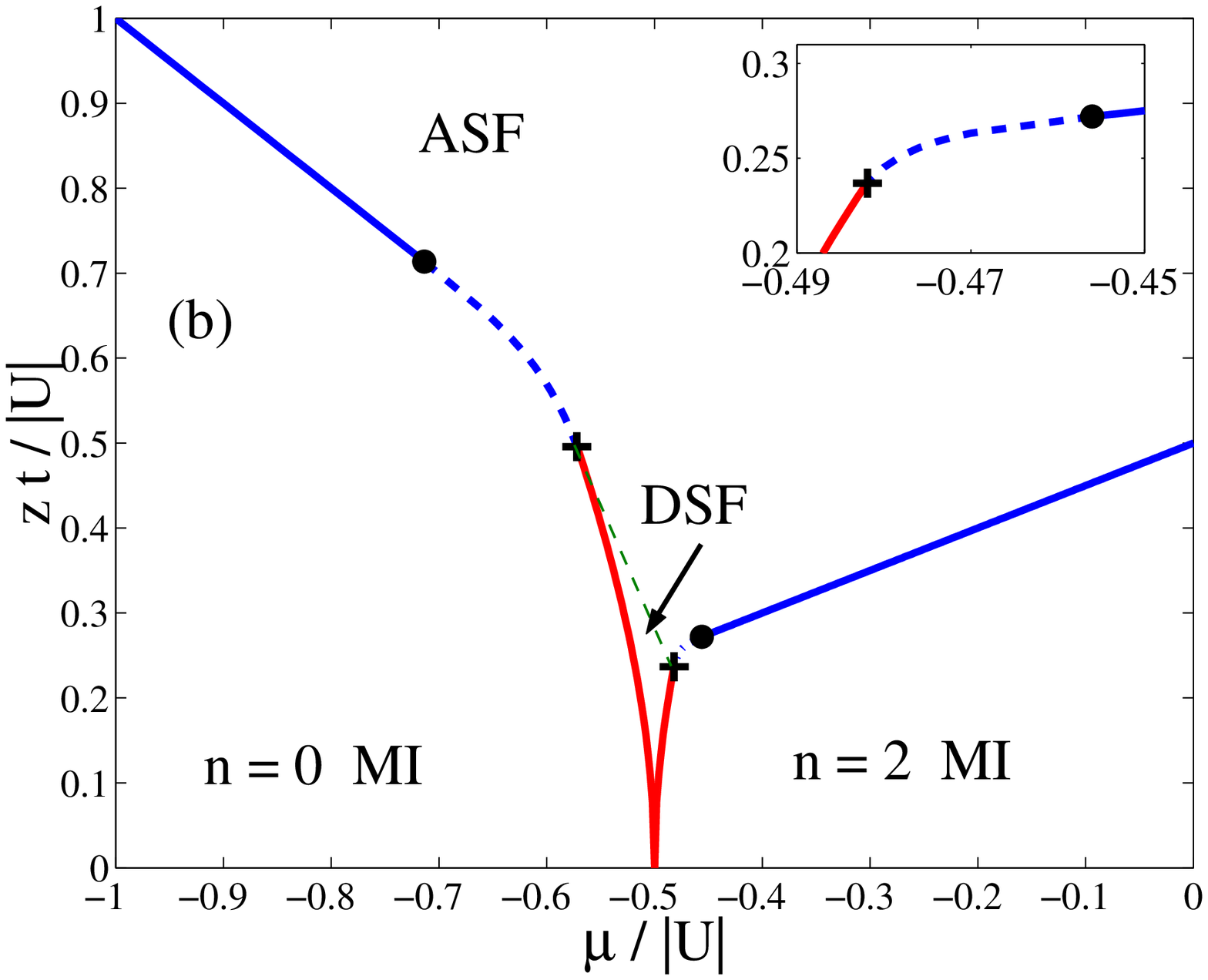}
\caption{(Color online) Phase diagram of the three-body
constrained attractive Bose-Hubbard model on (a) two-dimensional
square lattices (b) three dimensional cubic lattices. The red
solid lines indicate the continuous MI-DSF transitions, and the
blue solid (dashed) lines show the continuous (first-order) MI-ASF
transitions. The black dots (crosses) denote the tricritical
(critical end) points. The schematic phase boundary between the
ASF and the DSF phases is added as the green dashed line for
clarity. The insets show the details of the first-order $n=2$ MI-ASF 
transition lines.} \label{fig:phase_diag}
\end{figure}

As usual discussions for Ginzburg-Landau theory, the
\emph{continuous} MI-DSF and MI-ASF transitions occur when the
coefficients of the quadratic terms for the corresponding order
parameters vanish. That is, $m_2 = 0$ and $m_1 = 0$ give
continuous MI-DSF and MI-ASF transitions,
respectively~\cite{note}.
From the relation between $m_2$ and $r_2$ 
derived above and the perturbative results for $r_2$ in
Eqs.~\eqref{eq:coef-0MI} and \eqref{eq:coef-2MI}, the critical
value of hopping parameter for the continuous $n=0$ MI-DSF
transition is given by
$z\,\bar{t}_{c,\textrm{0MI-DSF}} \simeq
\sqrt{z\bar{\mu}(2\bar{\mu}+1) / 2}$,
while that for the continuous $n=2$ MI-DSF transition is
$z\,\bar{t}_{c,\textrm{2MI-DSF}} \simeq
\sqrt{z(\bar{\mu}+1)(2\bar{\mu}+1) / 2}$.
Similar reasoning applies also for the continuous MI-ASF
transition. Our calculations give the critical values of $\bar{t}$
for the continuous $n=0$ MI-ASF transition:
$z\,\bar{t}_{c,\textrm{0MI-ASF}} \simeq -\bar{\mu}$,
and that for the continuous $n=2$ MI-ASF transition:
$z\,\bar{t}_{c,\textrm{2MI-ASF}} \simeq (\bar{\mu} + 1) / 2$.
The phase boundaries of continuous MI-DSF and MI-ASF transitions
for bosons hopping on two-dimensional (2D) square lattices with
coordination number $z = 4$ are depicted in
Fig.~\ref{fig:phase_diag}(a) as the red and the blue solid lines,
respectively.

As mentioned before, the presence of the Feshbach resonance term
plays an important role in characterizing the nature of the ASF
state, and it may even modify the feature of the phase
transitions. In general, the phase boundaries of the MI phases are
determined by minimizing the effective potential $\Gamma$ in
Eq.~\eqref{eq:Gamma}, which leads to the following extremum
equations:
\begin{subequations}
\begin{align}
&\phi_{\rm a} ( m_1 + g_1 |\phi_{\rm a}|^2 - \alpha \phi_{\rm m} + g_{12} |\phi_{\rm m}|^2 ) = 0 \; , \label{eq:Saddle-1}\\
&-\frac{\alpha}{2} \phi_{\rm a}^2 %
 + \phi_{\rm m} ( m_2 + g_2 |\phi_{\rm m}|^2 + g_{12} |\phi_{\rm a}|^2 ) = 0 \; . \label{eq:Saddle-2}%
\end{align}
\end{subequations}
From Eq.~\eqref{eq:Saddle-2}, it is clear that, once $\phi_{\rm
a}$ becomes nonzero, the molecular condensate $\phi_{\rm m}$ will
acquire a non-vanishing value also. Thus the fact that the atomic
condensate and the molecular condensate are both finite in the ASF
phase can be explained straightforwardly as long as the Feshbach
resonance term exists in $\Gamma$. Nevertheless, since $\phi_{\rm
a} = 0$ in both the DSF and the MI phases, one can realize again
from Eq.~\eqref{eq:Saddle-2} that the condition for the continuous
MI-DSF transition is not affected by the presence of this
Feshbach-resonance term. Similarly, far away from the ASF-DSF
phase boundary, $\phi_{\rm m} \simeq 0$ in the ASF phase and
$\phi_{\rm m} = 0$ in the MI phase, thus the effect of the
Feshbach resonance coupling on the aforementioned continuous
MI-ASF transition should be negligible, which can be understood
from Eq.~\eqref{eq:Saddle-1}.


However, around the parameter region such that both $m_1$ and
$m_2$ are zero, the continuous MI-to-superfluid transitions
mentioned above will be preempted by \emph{first-order} MI-ASF
transitions caused by the Feshbach resonance term. Therefore,
crossing this phase boundary, the atomic and the molecular
condensates as well as the particle density will no longer change
continuously. The occurrence of this first-order MI-ASF transition
seems to be overlooked in
Refs.~\cite{Diehl09-1,Diehl09-2}.
As explained below, this
conclusion can be reached by carefully analyzing the extremum
equations in Eq.~\eqref{eq:Saddle-1} and \eqref{eq:Saddle-2},
which has been discussed in the context of the boson Feshbach
resonance problem~\cite{1QPT}.

Without loss of generality, we may take both $\phi_{\rm a}$ and
$\phi_{\rm m}$ to be real.
We remind that, on the first-order MI-ASF phase boundary, there
should exist a discrete jump for $\phi_{\rm a}$ to a finite value.
Therefore, to determine this first-order phase boundary, we need
to seek for nontrivial solutions of $\phi_{\rm a}$. From
Eq.~\eqref{eq:Saddle-1}, this is given by $|\phi_{\rm a}|^2 = (
\alpha \phi_{\rm m} - g_{12} |\phi_{\rm m}|^2 -m_1) / g_1$.
Substituting this condition into Eqs.~\eqref{eq:Saddle-2}, one
leads to a saddle-point equation, which is expressed solely in
terms of $\phi_{\rm m}$,
\begin{align}
&\frac{m_1\alpha}{2g_1} %
 + \left( m_2 - \frac{m_1g_{12}}{g_1} - \frac{\alpha^2}{2g_1} \right) \phi_{\rm m} %
 + \frac{3\alpha g_{12}}{2g_1} \phi_{\rm m}^2 \nonumber\\
&+ \left( g_2 - \frac{g_{12}^2}{g_1} \right) \phi_{\rm m}^3 = 0 \; .\label{eq:Saddle-3}%
\end{align}
Besides, the corresponding energy density difference between the
ASF and the MI phases is given by
\begin{align}
\Delta \Gamma &= -\frac{m_1^2}{2g_1} + \frac{m_1\alpha}{g_1} \phi_{\rm m} %
 + \left( m_2 - \frac{m_1 g_{12}}{g_1} - \frac{\alpha^2}{2g_1} \right) \phi_{\rm m}^2 \nonumber\\
&+\frac{\alpha g_{12}}{g_1} \phi_{\rm m}^3 %
 + \left( \frac{g_2}{2} - \frac{g_{12}^2}{2g_1} \right) \phi_{\rm m}^4 \; .\label{eq:ASF-Mott} %
\end{align}
The first-order MI-ASF phase boundaries are determined by the
simultaneous solutions of Eq.~\eqref{eq:Saddle-3} and $\Delta
\Gamma = 0$ in Eq.~\eqref{eq:ASF-Mott}. Moreover, the tricritical
points separating the continuous and the first-order MI-ASF
transitions can be found by satisfying the conditions for both
kinds of transitions. That is, by requiring nontrivial solution of
Eq.~\eqref{eq:Saddle-3} with $m_1 = 0$, one leads to the condition
for the tricritical points:
$m_2 - \alpha^2 / 2g_1 = 0$.
From the perturbative results for the coefficients of the free
energy density $f$ in Eqs.~\eqref{eq:coef-0MI} and
\eqref{eq:coef-2MI} as well as
the relations 
between the coefficients in $f$ and those in the effective
potential $\Gamma$, these first-order transition lines and the
corresponding tricritical points can be calculated numerically.
Our results of the first-order MI-ASF phase boundaries for the
models defined on 2D square lattices with $z = 4$ are depicted as
the blue dashed lines in Fig.~\ref{fig:phase_diag}(a). The value
of the tricritical point on the $n = 0$ MI-ASF transition line is
found to be $(\bar{t}_{\rm T}, \bar{\mu}_{\rm T}) \simeq (0.226,
-0.903)$, while $(\bar{t}_{\rm T}, \bar{\mu}_{\rm T}) \simeq
(0.071, -0.430)$ for that on the $n = 2$ MI-ASF transition line.
On the other hand, the value of the critical end point of the $n =
0$ MI-DSF transition line is given by $(\bar{t}_{\rm E},
\bar{\mu}_{\rm E}) \simeq (0.138, -0.624)$, while $(\bar{t}_{\rm
E}, \bar{\mu}_{\rm E}) \simeq (0.061, -0.472)$ for that of the $n
= 2$ MI-DSF transition line.
These tricritical (critical end) points are denoted by black dots
(crosses) in Fig.~\ref{fig:phase_diag}(a).

Within the present approach, the dependence of the lattice
structure and the dimensionality enter only through the
coordination number $z$. As an illustration for its influence, the
results for the models on three-dimensional cubic lattice with $z
= 6$ are presented in Fig.~\ref{fig:phase_diag}(b). We find that
the ranges in $\bar{\mu}$ (and in $\bar{t}$ also) for both the
first-order transition lines and the DSF phase are narrower than
those in the 2D case~\cite{note2}. This implies that the role
played by the Feshbach resonance term becomes less important as
$z$ increases.


To summarize, the phase boundaries of the MI-ASF and the MI-DSF
transitions in the three-body constrained attractive Bose lattice
gas are determined by generalizing the effective potential approach
developed in Refs.~\cite{dosSantos09,Bradlyn09}. While the
coefficients of the free energy density $f$ are calculated only up
to leading order in $t/|U|$, our results should be quantitatively
accurate at least in the strong-coupling region of the phase
diagram. We find that, due to the presence of the Feshbach
resonance term in the Ginzburg-Landau expansion, the continuous
MI-to-superfluid transitions can be driven to be
\emph{first-order} ones. From our results, it is found that the
DSF phase exists only in a narrow region of chemical potential
$\mu/|U|$ for small hopping parameters $t/|U|$. Therefore,
carefully tuning system parameters into the suggested parameter
regime are necessary to uncover experimentally this novel phase in
real ultracold Bose gas in optical lattices.
%

We are grateful to K.-K. Ng for for many enlightening discussions.
Y.-W. Lee and M.-F.Yang thank the support from the National
Science Council of Taiwan under grant NSC 96-2112-M-029-006-MY3
and NSC 96-2112-M-029-004-MY3, respectively.


\begin{thebibliography}{32}
\bibitem{note2}
Now the value of the tricritical point on the $n = 0$ MI-ASF
transition line becomes $(\bar{t}_{\rm T}, \bar{\mu}_{\rm T})
\simeq (0.119, -0.713)$, while $(\bar{t}_{\rm T}, \bar{\mu}_{\rm
T}) \simeq (0.045, -0.456)$ for that on the $n = 2$ MI-ASF
transition line. On the other hand, the value of the critical end
point of the $n = 0$ MI-DSF transition line is given by
$(\bar{t}_{\rm E}, \bar{\mu}_{\rm E}) \simeq (0.083, -0.572)$,
while $(\bar{t}_{\rm E}, \bar{\mu}_{\rm E}) \simeq (0.039,
-0.482)$ for that of the $n = 2$ MI-DSF transition line.
\end{thebibliography}
\end{document}